\documentclass{aa}  
\usepackage{comment}

\usepackage{graphicx}
\usepackage{txfonts}
\usepackage{lipsum}
\usepackage{subcaption}       
\usepackage{lscape}             
\usepackage{placeins}       
\usepackage[T1]{fontenc}                   

\usepackage[colorlinks,linkcolor=blue,citecolor=blue,urlcolor=blue,bookmarks=false,hypertexnames=true]{hyperref}
\usepackage{natbib}
\bibpunct{(}{)}{;}{a}{}{,}

\begin{document}

   \title{A Monte Carlo method for tracking dust properties during coagulation in protoplanetary disks}

   \author{Nerea Gurrutxaga\inst{1}
        \and Vignesh Vaikundaraman \inst{1} \and Joanna Dr{\k{a}}{\.z}kowska \inst{1}
        }

   \institute{Max Planck Institute for Solar System Research, Justus-von-Liebig-Weg 3, 37077 Göttingen, Germany\\
             \email{gurrutxaga@mps.mpg.de}
             \\ }

   \date{Received February 20, 2026; accepted April 7, 2026}

  \abstract
   {Dust growth is a crucial step in planet formation, and the efficiency of this process is controlled by the physical and chemical properties of the dust grains. Monte Carlo–based methods are commonly used to follow the collisional evolution of dust while tracking their properties. However, current Monte Carlo methods in planet formation do not strictly conserve the global inventory of dust properties across the protoplanetary disk, causing fluctuations that can grow over time and affect predictions of dust evolution. Here we present a coagulation algorithm that ensures the global conservation of dust properties while resolving the spatial evolution of dust. The method is validated against analytical solutions for standard coagulation kernels and benchmarked in a two-dimensional disk. We show that the method reproduces standard results, resolves the full dust population, and improves the resolution of the small-grain regime compared to other Monte Carlo methods for modeling global dust evolution. Finally, using a test case that includes sublimation and condensation of water interacting with silicates, we demonstrate strict conservation of each component's mass during coagulation, establishing the method as a valuable tool for tracking dust properties in protoplanetary disks.}


   \maketitle
   \nolinenumbers

\section{Introduction}

   Planet formation requires dust particles to grow across many orders of magnitude in size within protoplanetary disks. Micrometer-sized dust coagulates into centimeter-sized pebbles before it can form kilometer-sized planetesimals \citep{Youdin2005} or be accreted by growing protoplanets \citep{Ormel2010, Lambrechts2012}. The physical and chemical properties of dust particles affect their transport and collision outcomes and thereby the efficiency of dust coagulation \citep{Birnstiel2024}. A comprehensive understanding of planet formation, therefore, requires studying how dust properties affect coagulation. This task is challenging, as resolving dust coagulation in detail is computationally demanding.

    Two main approaches are commonly used to model dust coagulation in protoplanetary disks. The first class consists of Monte Carlo methods that explicitly track particle evolution. These methods were introduced to planet formation studies independently by \cite{Ormel2007} and \cite{Zsom2008}, with later developments by \cite{Ormel2008, Drazkowska2013} and \cite{Beutel2023}. The implementation described in \cite{Drazkowska2013} is available as the open-source code \texttt{mcdust} \citep{Vaikundaraman2025}. The second widely used approach is the mass grid-based method \citep[e.g.,][]{Weidenschilling1980, Dullemond2005, Brauer2008, Birnstiel2010}, in which the dust mass distribution is discretized into bins and interactions between bins are calculated. This approach is well established in the community, with open-source implementations such as the one-dimensional code \texttt{dustpy} \citep{Stammler2022} and the two-dimensional code \texttt{cuDisc} \citep{Robinson2024}. A major challenge of the mass-grid method is the inclusion of additional particle properties, as they substantially increase the complexity and computational cost of the method. While extensions to account for properties such as porosity, chemical composition, and charge are possible \citep{Ossenkopf1993, Okuzumi2009, Stammler2017, Akimkin2023}, they require simplifying assumptions -- such as that dust particles of equal mass share the same properties. In contrast, Monte Carlo methods enable a more straightforward implementation of additional properties without such assumptions, although adequately sampling the larger parameter space still requires increased numerical resolution.

    Monte Carlo methods widely used in planet formation studies are based on the stochastic description of coagulation introduced by \cite{Gillespie1975}. Because it is not feasible to track all dust particles in a protoplanetary disk, these methods rely on a limited number of representative particles, whose evolution is assumed to describe that of many physical particles. There are different ways to define and evolve such representative particles, each way leading to different strengths and limitations. In the approach presented by \cite{Ormel2007}, collisions are computed between individual representative particles confined within a small volume that represents a larger volume. When two particles coagulate, a new representative particle is created by randomly duplicating an existing one in order to keep the total number of representative particles constant. To preserve the dust density, the volume enclosing the particles is increased accordingly. This method can achieve high accuracy, particularly during the earliest stages of growth. An improved and more efficient implementation was presented by \cite{Ormel2008}, showing that several physical collisions can be grouped into a single collision event. However, the general approach by \cite{Ormel2007} is designed to resolve dust coagulation locally, which can make it challenging to study global dust evolution, as dust transport influences coagulation across the disk \citep{Birnstiel2012}. An alternative Monte Carlo method was introduced by \cite{Zsom2008}. In this approach, a small number of representative particles is tracked, each representing a large group of identical physical particles. To compute collisions, the representative particle $i$ is assumed to collide with physical particles identical to those represented by particle $j$. However, only particle $i$ is updated after the collision. It was shown that, when a sufficient number of representative particles is used, the expected coagulation behavior is well reproduced. This method has two key advantages: it allows for spatially resolved dust coagulation, and it requires a relatively small number of particles to capture the statistical evolution of the system \citep{Drazkowska2013}.
    
    Monte Carlo methods facilitate the inclusion of additional particle properties with minimal modification to the algorithm and have already been employed to study porosity \citep{Ormel2007, Zsom2008, Krijt2015} and chemical composition \citep{Krijt2016, Houge2023, Vaikundaraman2025_master}. However, \cite{Krijt2016} noted that although the approach by \cite{Zsom2008} can conserve the total dust mass, it does not strictly conserve the mass of different components. This arises because of the asymmetry of only updating the representative particle $i$ during a collision with a physical particle $j$, thereby introducing fluctuations in the composition. These fluctuations become particularly significant when size-dependent processes operate and colliding dust particles have largely different compositions. In the alternative method by \cite{Ormel2007}, the global inventory of dust properties is conserved during pairwise collisions. However, since a particle is duplicated to keep the number of representative particles constant, fluctuations can also arise when dust particles have largely different compositions.

    A Monte Carlo method that both strictly conserves quantities such as dust composition and enables spatial resolution is still lacking. This limitation can hinder the development of more advanced models. In this work, we present a Monte Carlo approach that is spatially resolvable and enforces strict conservation of the global inventory of dust properties. Our method is comparable to that presented by \cite{Shima2009} for cloud microphysics but is formulated here in the context of planet formation. In Sect.~\ref{sec:algorithm}, we describe the algorithm. In Sect.~\ref{sec:tests}, we validate the method through a series of tests by comparing our results with analytical solutions of the Smoluchowski equation for coagulation kernels, computing dust evolution in a two-dimensional protoplanetary disk, and following the collisional evolution of a multi-component dust population. We summarize our findings in Sect.~\ref{sec:conclusion}.

\begin{figure}
    \centering
    \includegraphics[width=8cm]{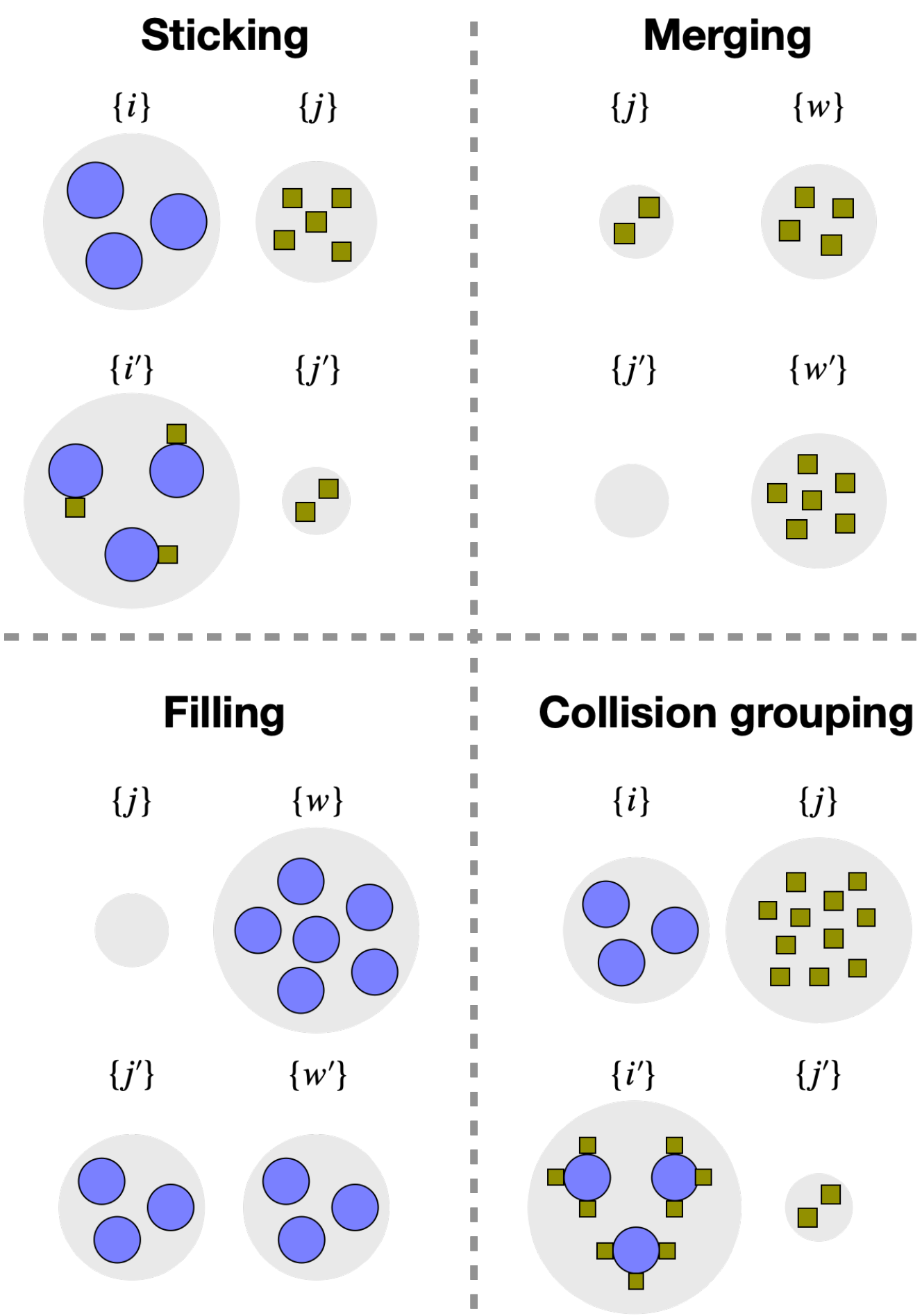}
    \caption{Summary of the Monte Carlo algorithm presented in this paper. We group individual identical particles. \textit{Sticking}: When group $\{ i\}$ collides with group $\{ j\}$, as $N_{i}\,$$<\,$$N_{j}$, every particle in group $\{ i\}$ sticks to one particle that belongs to group $\{ j\}$. \textit{Merging}: If group $\{ j\}$ contains negligible mass, it merges with group $\{ w\}$, with $j$ and $w$ being particles with similar characteristics. \textit{Filling}: When group $\{ j\}$ is emptied, it is filled with half of the particles from the largest group $\{ w\}$. \textit{Collision grouping}: If group $\{ i\}$ collides with $\{ j\}$, with particle $j$ being very small compared to particle $i$, we group collisional events, meaning that in a single collision we stick several $j$ particles into a single particle $i$.}
     \label{fig:schematic}
\end{figure}
\section{The Monte Carlo algorithm}\label{sec:algorithm}
We assume that the total population of $N_{\rm{tot}}$ dust particles can be represented by $N_{r}$ representative particles. A representative particle $i$ represents the evolution of $N_{i}$ identical particles of mass $m_{i}$ that constitute the group $\{i\}$ with a total group mass of $M_{i}\,$$=\,$$N_{i} m_{i}$. With $M_{\rm{tot}}$ being the total population mass,
\begin{eqnarray}
\label{eq:totpar}
  N_{\rm{tot}} & = &\sum_{i=1}^{N_{r}}N_{i}\,,\\
 \label{eq:totmass}
 M_{\rm{tot}} & = &\sum_{i=1}^{N_{r}}M_{i}\,.
\end{eqnarray}
$N_{\rm{tot}}$ decreases over time as particles coagulate into larger particles. However, $M_{\rm{tot}}$ should be conserved over time.

In the model presented here, we group particles that undergo one-to-one collisions in parallel as independent Poisson processes. A similar strategy was adopted by other authors \citep{Ormel2008, Zsom2008, Shima2009}, although the grouping procedure differs between them. In our approach, as in \cite{Zsom2008} and \cite{Shima2009}, the number of representative particles, $N_{ r}$, equals the number of particle groups and remains constant throughout the simulation. In contrast, in \cite{Ormel2008} the number of representative particles (or species) and groups are not equal, and either or both can vary depending on the specific method. When two groups $\{i\}$ and $\{j\}$ collide, we update particles involved in the collision from both groups as in \cite{Ormel2008} and \cite{Shima2009}. By contrast, in the method of \cite{Zsom2008}, only one group is updated per collision event. First, in Sect.~\ref{sec:colrates} we explain how to calculate the collision rate and choose collision events between pairs of particle groups. In Sect.~\ref{sec:sticking}, we describe how sticking collisions between two groups of particles are modeled. In Sect.~\ref{sec:filling} and \ref{sec:grouping}, we discuss details of the algorithm that improve accuracy and computational efficiency. Figure \ref{fig:schematic} summarizes the algorithm.

\subsection{Collisions between groups of particles} \label{sec:colrates}

The collision rate $C_{i,j}$ between particle $i$ and $j$ in a volume $V$ that encloses $N_{i}$ and $N_{j}$ particles of each group is given by
\begin{equation}\label{eq:colrate}
    C_{i,j} = N_{i}N_{j}\frac{K_{i,j}}{V}\,,
\end{equation}
where $K_{i,j}$ is a coagulation kernel. In protoplanetary disks, the kernel is typically defined as
\begin{equation}\label{eq:kernel}
    K_{i,j} = \Delta v_{i,j}\sigma_{i,j}\,,
\end{equation}
where $\Delta v_{i,j}$ is the relative velocity between particle $i$ and $j$ and $\sigma_{i,j}\,$$=\,$$\pi(a_{i}+a_{j})^{2}$ is the geometrical cross section of their collision, with $a_{i}$ and $a_{j}$ denoting the particle radii.

The number of collisions required for all particles from group $\{i\}$ or $\{j\}$ to collide is given by $N_{\rm{col}}\,$$ =\,$$ \min{(N_{i}, N_{j})}$. When one-to-one collisions are grouped, the effective collision rate between representative particles $i$ and $j$ is
\begin{equation}\label{eq:group_colrate}
    \tilde{C}_{i,j} = \frac{C_{i,j}}{N_{\rm{col}}} = \frac{N_{i}N_{j}\frac{K_{i,j}}{V}}{\min(N_{i}, N_{j})}\,.
\end{equation}
If $N_{i}\,$$<\,$$N_{j}$, a grouped collision event takes place when every particle in group $\{i\}$ collides with a particle in group $\{j\}$. In that case, the collision rate simplifies to
\begin{equation}\label{eq:colrate_Ni}
    \tilde{C}_{i,j} =N_{j}\frac{K_{i,j}}{V}\,,
\end{equation} 
which corresponds to the collision rate between a defined grouped $\{i\}$ to collide with $N_{j}$ particles \cite[][their Eq.~1]{Liffman1992}. The grouped collision rate in Eq.~\ref{eq:group_colrate} is symmetric, such that $\tilde{C}_{i,j}\,$$=\,$$\tilde{C}_{j,i}$. This differs from the method described by \cite{Zsom2008}.

Once the collision rate between groups is determined, we compute the collision probabilities using the full-conditioning method of \cite{Gillespie1975}. The collision probability density function $P(\tau,i,j)$ can be expressed as the product of three one-variable probability functions:
\begin{equation}\label{eq:full-condition}
    P(\tau,i,j) = P_{1}(\tau) \cdot P_{2}(i \mid \tau) \cdot P_{3}(j \mid i)  \,.
\end{equation}
Here, $P_{1}(\tau)$ is the probability that the next collision occurs within the time interval $[t, t+\tau]$, $P_{2}(i \mid \tau)$ is the probability that group $\{i\}$ participates in the given collision, and $P_{3}(j \mid i)$ is the probability that group $\{j\}$ participates in the collision, knowing that group $\{i\}$ is involved. Defining the total collision rate as $C_{\rm tot}$ and the collision rate of group $\{i\}$ as $\tilde{C}_{i}$,
\begin{eqnarray}
\label{eq:total}
  C_{\rm tot} &= &\sum_{i=1}^{N_{r}} \tilde{C}_{i} \,,\\
  \label{eq:Ci}
\tilde{C}_{i} &= &\sum_{j=i}^{N_{r}} \tilde{C}_{i,j}\,.
\end{eqnarray}
We include the possibility that group $\{i\}$ collides with itself. In our method, this can be allowed because half of the particles in $\{i\}$ can collide with the other half. When computing the collision rates in Eq.~\ref{eq:group_colrate}, we account for this by replacing $N_{i}$ with $N_{i}/2$. Consequently, there are $N_{r}(N_{r}+1)/2$ distinct collision pairs.

Each probability in Eq.~\ref{eq:full-condition} is defined as
\begin{eqnarray}
\label{eq:probability}
  P_{1}(\tau) &=& C_{\rm tot} \exp{(-C_{\rm tot}}\tau), \quad 0\leq\tau <\infty\,,\\
P_{2}(i \mid \tau) &=& \frac{\tilde{C}_{i}}{C_{\rm tot}} \,,\\
P_{3}(j \mid i) &=& \frac{\tilde{C}_{i,j}}{\tilde{C}_{i}}\,.
\end{eqnarray}
By generating three random numbers from a uniform distribution, we determine both the time of the next collision and the pair of groups involved. 
\subsection{Sticking}\label{sec:sticking}

When two groups collide, the properties of groups $\{i\}$ and $\{j\}$ are modified according to the collision outcome, which depends on their relative velocity $\Delta v_{ij}$. Above specific velocity thresholds, collisions result in bouncing, fragmentation, or erosion, while below these thresholds, particles stick \citep[e.g., ][]{Blum1993, Guttler2010, Zsom2010, Schrapler2011, Kelling2014}. Here we describe how sticking events are computed. For clarity, we assume $N_i\,$$\leq\,$$ N_j$ in this section; the opposite case is handled by exchanging the indices.

When particles in groups $\{i\}$ and $\{j\}$ collide at velocities that enable sticking, every particle in group $\{i\}$ coagulates with a particle in group $\{j\}$ (see Fig.~\ref{fig:schematic}). In a single simulated collision event, the $N_{i}\,$$\times\,$$m_{j}$ mass is transferred from group $\{j\}$ to group $\{i\}$. We update the mass of representative particles $i$ and $j$, as well as the number of particles $N_{i}$ and $N_{j}$ constituting groups $\{i\}$ and $\{j\}$ as
\begin{eqnarray}
\label{eq:stick}
 m^{\prime}_{i} &=& m_{i} + m_{j}\,,\\
m^{\prime}_{j}&=&m_{j}\,, \\
N^{\prime}_{i}&=& N_{i}\,, \\
\label{eq:last}
N^{\prime}_{j}&=&N_{j}-N_{i}\,.
\end{eqnarray}
After each sticking event, the collision rates in Eq.~\ref{eq:colrate_Ni} for every pair involving groups $\{i\}$ and $\{j\}$ are recalculated, and Eqs.~\ref{eq:total} and \ref{eq:Ci} are updated accordingly.

If $N_{i}\,$$=\,$$ N_{j}$, group $\{j\}$ is emptied with $N^{\prime}_{j} = 0$ when sticking. The empty group $\{j\}$ can then be reused to split another group $\{w\}$, thereby better resolving the evolution of particles of type $w$. Our criterion for splitting groups is described in Sect.~\ref{sec:filling}.

In this method, we advance the system over multiple independent collisions in a single step. The approximation is valid provided that the system properties do not vary significantly over this sequence. This condition is most easily satisfied during sticking events involving particles with different mass, since coagulation between similar particles produces the largest relative changes in the system. The approximation is also more accurate for collisions involving the largest particles, since a group of large particles typically contains fewer particles. In Sect.~\ref{sec:tests}, we show that grouping one-to-one collisions provides reliable estimates for typical dust evolution problems.

\subsection{Merging and filling the empty group}\label{sec:filling}

When a group $\{j\}$ becomes empty, the freed memory can be used to split an existing group $\{w\}$ into two (see Fig.~\ref{fig:schematic}). Unlike the approaches in \cite{Ormel2007} and \cite{Ormel2008}, we did not duplicate particles. Instead, we began from large particle groups and increased the resolution by splitting existing groups at later times, as similarly done by \cite{Shima2009}.

A convenient choice of which group $\{w\}$ to split improves the resolution, whereas a poor choice generates particles that occupy space but are unimportant to the system’s evolution, as they represent only a tiny fraction of the total mass. In a coagulation event between group $\{i\}$ and $\{j\}$, with $N_{i}\,$$\leq\,$$N_{j}$, group $\{i\}$ gains mass, while group $\{j\}$ losses mass. If the collision probability increases with particle mass, as in protoplanetary disks, group $\{i\}$ is expected to play an increasingly important role in the subsequent collisional evolution. It is therefore more likely to undergo further collisions and continue growing. When a single group $\{i\}$ dominates the mass distribution while still comprising many particles, the grouped description becomes inaccurate, because it artificially reduces the particles’ growth rate compared to an ungrouped treatment. A practical way to mitigate this effect is to preferentially split the group $\{i\}$ that contains the largest group mass $M_{i}\,$$=\,$$N_{i}\,$$m_{i}$ (see Fig.~\ref{fig:schematic}). Splitting the groups with the largest mass increases the resolution in the high-mass regime. Alternative splitting criteria can be employed to resolve the distribution of smaller particles better. For instance, one may adopt an approach similar to \cite{Ormel2008}, dividing the size distribution into exponentially spaced mass bins while aiming to maintain a similar number of representative particles per bin.

Even with an appropriate choice of how groups are filled, some groups may still contribute negligibly to the collisional evolution of the system. This occurs because group $\{j\}$ typically requires several coagulation steps to lose all its mass. To accelerate this process, we introduced a group-mass threshold $M_{\rm min}$, and groups with masses below this value are considered negligible and are emptied. We define
\begin{equation}\label{eq:merging}
    M_{\rm min} = X\cdot \frac{M_{\rm tot}}{N_{r}}\,,
\end{equation}
where $M_{\rm tot}/N_{r}$ is the average mass of a group (see Eq.~\ref{eq:totmass}), and $X$ is a user-defined merging parameter ($0\,$$\leq\,$$X<\,$$1$). A lower value of $X$ leads to fewer merging events and, consequently, fewer large groups are split. Therefore, achieving higher resolution among the largest particles requires choosing a higher value of $X$. Conversely, if the focus is on the small-particle population, $X$ can be reduced (see further discussion in Sect.~\ref{sec:global}).

When group $\{j\}$ is forced to empty, it can be merged with its most similar group $\{w\}$ to minimize the noise added to the simulation \citep{Ormel2008}. We quantify similarity between groups by calculating the distances in the property space. In Sects.~\ref{sec:kernel} and \ref{sec:global}, we consider only the particle mass $m_i$ as a relevant property for merging, and therefore the distance between groups $\{i\}$ and $\{j\}$ is defined as
\begin{equation}\label{eq:distance_mass}
d_{ij} = |m_i - m_j|\,.
\end{equation}
We compute $d_{ij}$ for all $i\,$$\neq\,$$j$ and find the group $\{w\}$, which has the shortest distance $d_{wj}$. In Sect.~\ref{sec:water}, we track water and silicate masses, $m_{\mathrm{ice},i}$ and $m_{\mathrm{si},i}$, and thus define distance as
\begin{equation}\label{eq:similar_water}
d_{ij} = \sqrt{
(m_{\mathrm{ice},i} - m_{\mathrm{ice},j})^{\,2} +
(m_{\mathrm{si},i} - m_{\mathrm{si},j})^{\,2}
}\,.
\end{equation}
Equation \ref{eq:similar_water} is especially suitable for the multicomponent simulation discussed in Sect.~\ref{sec:water} because we can express properties in terms of mass. However, there are instances where this is not possible, and one property may have a significantly broader range of values than others. In such cases, alternative merging criteria may be more suitable \citep[e.g.,][for particle mass and porosity]{Krijt2015}.

We treat merging as a process distinct from coagulation (see Fig.~\ref{fig:schematic}): when group $\{j\}$ merges with group $\{w\}$, their total group mass is combined, while the new mass of the representative particle $w$ is computed as an average such that
\begin{eqnarray}
\label{eq:merge}
m^{\prime}_{w} &=& \frac{N_{w}m_{w} + N_{j}m_{j}}{N_{w}+N_{j}}\,,  \\
m^{\prime}_{j}&=&0\,, \\
N^{\prime}_{w}&=& N_{w}+N_{j}\,, \\
N^{\prime}_{j}&=&0\,,
\end{eqnarray}
where $\{w\}$ is the group whose particle properties are most similar to those in group $\{j\}$. After group $\{j\}$ is emptied, it is refilled with half of the particles in the group $\{i\}$ that contains the largest group mass $M_{i}\,$. After each merging and filling event, the collision rates in Eq.~\ref{eq:colrate_Ni} of pairs involving any modified groups are recalculated, and Eqs.~\ref{eq:total} and \ref{eq:Ci} are updated accordingly.

As particles merge, their properties average out, losing information about the dispersion of those properties. To test the validity of the results against this artificial smoothing, we can vary the merging parameter $X$. If in simulations with little or no merging (i.e., low $X$), particle properties are more spread than in simulations with frequent merging events, this suggests that merging artificially smooths the distribution of particle properties. In such cases, it may be necessary to increase the number of representative particles $N_{r}$, reduce $X$, or modify the criteria for selecting the most similar particle.

\subsection{Physical collision grouping} \label{sec:grouping}
When the particle size distribution is broad, collisions between particles $i$ and $j$ with very different masses can be frequent. These collisions, however, change the mass of the largest particle only marginally in a single collisional event. It is therefore possible to combine several physical collisions into a single effective collision. 

\cite{Zsom2008} and \cite{Ormel2008} introduced numerical optimizations to speed up the calculation of this type of collision, which is also applicable to our method. In their approach, collisions between individual particles are grouped such that a single representative particle undergoes many physical collisions. It is important to note that this differs from the particle grouping described so far. Until now, we have considered the grouping of $N_{i}$ events across $N_{i}$ different particles, whereas in this section we focus on each particle experiencing multiple collisions (see Fig.~\ref{fig:schematic}).

\cite{Zsom2008} suggested grouping collisions based on the particle mass ratio, using a grouping minimum mass $dm_{\mathrm{max}}$, such that collisions are grouped when $m_j\,$$/\,$$m_i\,$$\leq\,$$dm_{\mathrm{max}}$. To implement this in our method, a sufficient number of physical particles must be available for collision grouping. If $m_j\,$$/\,$$m_i\,$$\leq\,$$dm_{\mathrm{max}}$, we modify the collision rate in Eq.~\ref{eq:colrate_Ni} as
\begin{equation}
    \tilde{C}^{\prime}_{i,j} = \frac{\tilde{C}_{i,j}}{n_{\rm group}}\,,
\end{equation}
where $n_{\rm group}$ is the number of physical collisions that every particle $i$ undergoes in one simulated event. The value of $n_{\rm group}$ depends on the mass ratio of particles and the number of physical particles: if $N_{i}\,$$/\,$$N_{j}\,$$\leq\,$$m_j\,$$/\,$$m_i\,$, $n_{\rm group }\,$$=\,$$(m_{i}/m_{j})\,dm_{max}$; otherwise, $n_{\rm group }\,$$=\,$$(N_{j}/N_{i})\,dm_{max}$.

\begin{figure*}
    \centering
    \includegraphics[width=15cm]{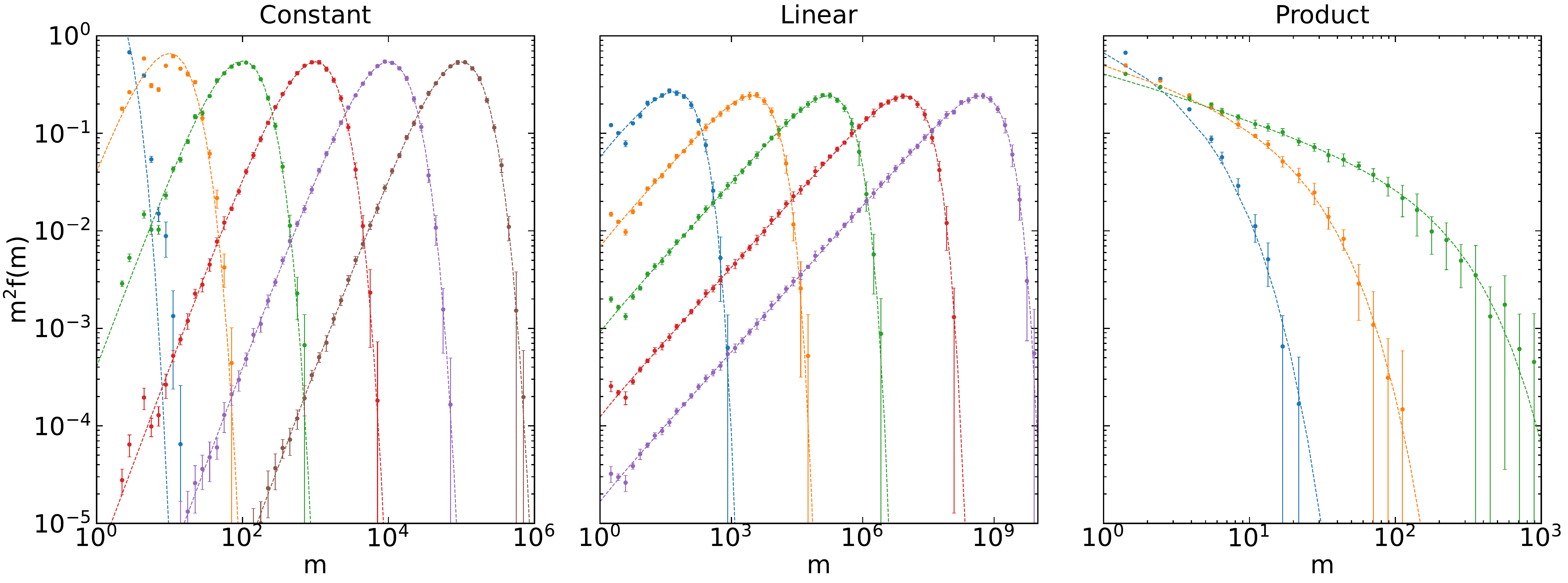}
    \caption{Particle mass distribution for the high-resolution tests against the analytical solutions of the Smoluchowski equation (dotted lines) evaluated at different dimensionless times. We used $10^{4}$ representative particles and the merging parameter $X\,$$=0.01$. We repeated each simulation ten times. Left: Constant kernel $K_{ij}\,$$=\,$$1$, evaluated at times $1$, $10$, $10^{2}$, $10^{3}$, $10^{4}$, and $10^{5}$. Center: Linear kernel $K_{ij}\,$$=\,$$\frac{1}{2}(m_i+m_j)$, evaluated at times $4$, $8$, $12$, $16$, and $20$. Right: Product kernel $K_{ij}\,$$=\,$$m_i\,$$ \times\,$$ m_j$, evaluated at times $0.4$, $0.7$, and $0.9$.}
     \label{fig:kernel}
\end{figure*}
\begin{figure*}
    \centering
    \includegraphics[width=15cm]{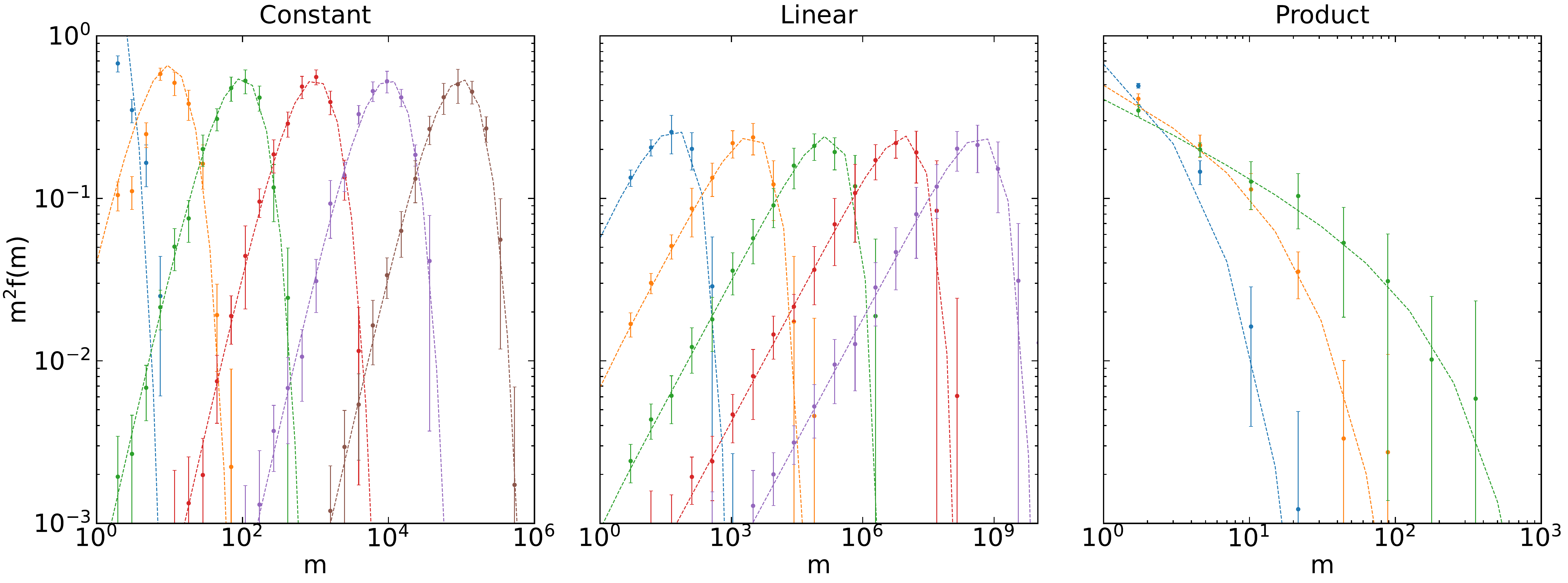}
    \caption{Particle mass distribution for the low-resolution tests against the analytical solutions of the Smoluchowski equation (dotted lines) evaluated at dimensionless times. We used $200$ representative particles and the merging parameter $X\,$$=\,$$0.1$. We repeated each simulation ten times and evaluated the same time instances, as in Fig.~\ref{fig:kernel}.}
     \label{fig:kernel_lowres}
\end{figure*}

We set $dm_{\mathrm{max}} = 10^{-2}$ for all tests in Sect.~\ref{sec:tests}. This value results in a $1\%$ error in particle mass, because when particle $i$ sticks to, for example, $n_{\rm group }\,$$=\,$$(m_{i}/m_{j})\,dm_{max}$ $j$-like particles, the particle mass $i$ is updated as 
\begin{equation}
    m^{\prime}_{i}=m_{i}+n_{\rm group}\,m_{j}=m_{i}+\left(\frac{m_{i}}{m_{j}}dm_{max}\right)m_{j}=(1+dm_{max})\, m_{i}\,.
\end{equation}

\section{Tests}\label{sec:tests}
\subsection{Analytical solutions of the Smoluchowski equation}\label{sec:kernel}

The Smoluchowski equation, commonly used to describe dust coagulation in astrophysics \citep{Smoluchowski1916}, has analytical solutions for three specific forms of the coagulation kernel: the constant kernel $K_{ij}\,$$=\,$$1$, the linear kernel $K_{ij}\,$$=\,$$\frac{1}{2}(m_i+m_j)$, and the product kernel $K_{ij}\,$$=\,$$m_i\,$$ \times\,$$ m_j$. Our coagulation model is required to reproduce these analytical solutions. We performed the analytical kernel tests as in \cite{Zsom2008} and \cite{Drazkowska2013}. We initialized all particles with a dimensionless mass $m_0\,$$=\,$$1$. We performed high- and low-resolution tests (see Figs.~\ref{fig:kernel} and \ref{fig:kernel_lowres}) and compared them against the analytical solutions given in Eqs.~(2.1-2.3) of \cite{Tanaka1994}.

For the high-resolution test, we used $10^{4}$ representative particles, each initially representing the same number of real particles. We adopted the merging parameter $X\,$$=\,$$0.01$ from Sect.~\ref{sec:filling} and repeated each simulation ten times. Figure \ref{fig:kernel} shows that in the high-resolution simulations, our algorithm accurately describes the evolution of the particle mass distribution over five orders of magnitude for the constant and linear kernels, and over four orders of magnitude for the product kernel. A notable feature seen in the constant kernel case is that the first two time steps are less accurate than the rest. This behavior arises because we assume that the system remains nearly unchanged over the time interval during which one-to-one collisions are grouped (see Sect.~\ref{sec:sticking}). Collisions between similar-sized particles produce the largest relative changes in the system, and the initial stages correspond to the largest number of real particles (see Eq.~\ref{eq:totpar}). As the mass distribution shifts toward larger masses, each group contains fewer particles, and fewer one-to-one collisions are grouped, which leads to improved accuracy. We conclude that this assumption has only a minor impact on the overall evolution of the system.
\begin{figure*}
    \centering
    \includegraphics[width=18cm]{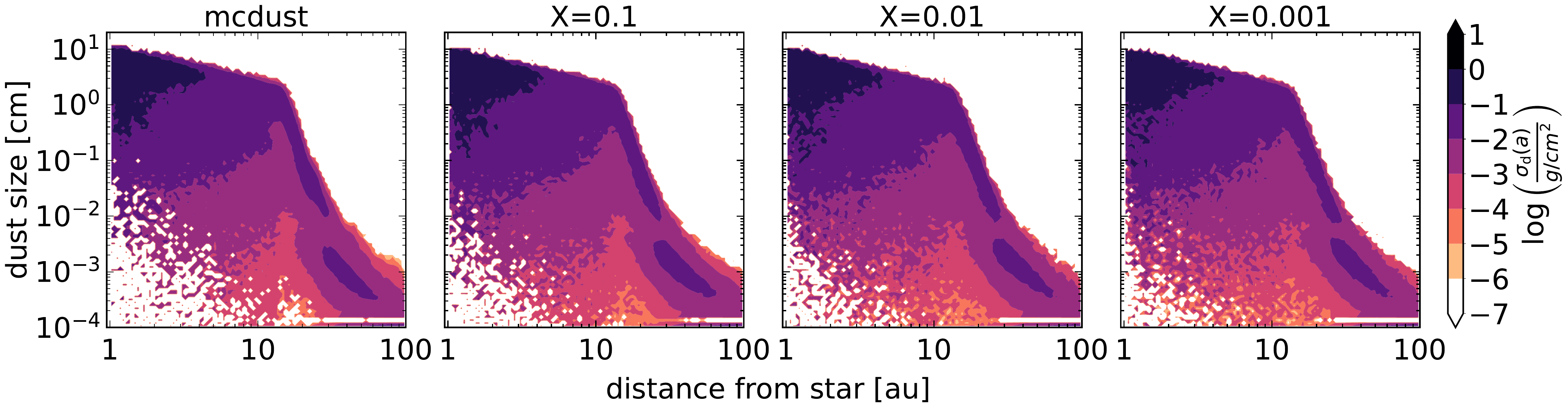}
    \caption{Dust size distribution in two-dimensional protoplanetary disks. The color map shows the logarithmic, size-dependent dust surface density $\sigma_{d}(a)$, computed with 100 radial and dust-size bins. Left panel: Reference simulation using the open-source code \texttt{mcdust}. Remaining panels: Results obtained with the algorithm presented in this work, adopting the merging parameters $X\,$$=\,$$0.1$, $0.01$, and $0.001$ (see Sect.~\ref{sec:filling}). For $X\,$$=\,$$0.01$ and $0.001$, the small-size range is more accurately resolved, whereas larger $X$ values reduce the resolution in this regime. The difference between $X\,$$=\,$$0.01$ and $0.001$ is minor, since a larger number of particles is required to fully benefit from values of $X\,$$<\,$$0.01$.}
     \label{fig:mcdust}
\end{figure*}

We then reduced the resolution to $200$ particles and increased the merging parameter to $X\,$$=\,$$0.1$ to assess the performance of the algorithm at low particle numbers. As in the previous test, the simulations were repeated ten times, and the results are presented in Fig.~\ref{fig:kernel_lowres}. As expected, the results are noisier than in the high-resolution simulation. However, we see that the algorithm can still describe the overall evolution for the three kernels. We find that below $200$ particles the linear and product kernels exhibit deviations from the analytical solutions. Therefore, we conclude that our algorithm requires at least $200$ representative particles, consistent with the findings of \cite{Drazkowska2013}. We further note that reducing the merging parameter $X$ below $0.01$ requires a larger number of particles to resolve the linear kernel accurately.

\subsection{Dust evolution in global protoplanetary disks} \label{sec:global}

In this section, we describe the ability of our method to capture dust evolution in a two-dimensional protoplanetary disk, including dust transport, growth, and fragmentation. To compute collisions between nearby particles, we used the adaptive radial–vertical grid described in \cite{Drazkowska2013}. We compared our algorithm with the publicly available code \texttt{mcdust} \citep{Vaikundaraman2025}, using the same prescriptions for dust transport and grid construction. The coagulation algorithm currently implemented in \texttt{mcdust} is based on \cite{Zsom2008} approach.

Overall, the \texttt{mcdust} algorithm works as follows. Dust particles are initialized and an initial grid is constructed with an equal number of representative particles per cell. Particle velocities are then computed, including radial drift, vertical settling, and radial and vertical turbulent diffusion (see Eqs.~7–18 in \citealt{Drazkowska2013}). The global time step $\Delta t_{\rm global}$ is set by the Courant condition, $\Delta t $$\,< $$\,\Delta x / v_{\rm max}$, ensuring that particles do not cross grid-cell boundaries within a single time step (see Sect.~2.5 of \citealt{Drazkowska2013}). The global timescale is further constrained by the collisional timescale by limiting the average number of collisions per particle, scaling as $\sim\,$$\Delta t^{\prime}_{\rm global}(n_{\rm cell}/n_{\rm coll})$, where $n_{\rm cell}$ and $n_{\rm coll}$ denote the number of particles and collisions per cell during the previous time step. Once $\Delta t_{\rm global}$ is determined, particles are transported over this time step. A new grid is then constructed from the updated positions, and collisions are computed within each cell. The time between consecutive collisions, $\tau$, is calculated via random sampling (see Eq.~\ref{eq:probability}) and accumulated until $\Delta t_{\rm global}$ is reached. The global time step is then updated and the procedure repeated.

For the algorithm presented here, we tested merging parameters of $X\,$$=\,$$0.1$, $0.01$, and $0.001$ in independent simulations, and Eqs.~\ref{eq:merging} and \ref{eq:distance_mass} were evaluated for each bin of the grid. Since each bin represents a single location in the disk, group positions remained unchanged during coagulation: when $\{j\}$ sticks to $\{i\}$, merges with $\{w\}$, or is filled with half of particles of $\{w\}$, the resulting groups $\{i'\}$, $\{w'\}$, and $\{j'\}$ remain at the original locations of $\{i\}$, $\{w\}$, and $\{j\}$, respectively.

In the current \texttt{mcdust}, when two representative particles collide at velocities above the fragmentation threshold, only one representative particle undergoes fragmentation. In reality, collision outcomes also depend on the mass ratio of particles \citep[e.g.,][]{Guttler2010, Hasegawa2021, Hasegawa2023}, but here we neglect it for simplicity. The new mass of the representative particle is chosen by randomly sampling from a fragment mass distribution $n(m)\,$$\propto\,$$ m^{-11/6}$ \citep{Birnstiel2011, Drazkowska2014}, and fragments can be composed of multiple monomers or continuous sections rather than being limited to discrete monomers. To construct a comparable model to \texttt{mcdust}, we assume that when groups $\{i\}$ and $\{j\}$ collide (with $N_i\,$$\,\leq$$\,N_j$), all particles in group $\{i\}$ fragment. Since there are $N_i$ real collision events, not all particles in group $\{j\}$ fragment in reality. In our method, however, all particles belonging to the same group evolve identically, so we assign a probability of $N_i\,$$/\,$$N_j$ that all particles in group $\{j\}$ fragment; otherwise, they remain intact. After a group fragments, the collision rates in Eq.~\ref{eq:colrate_Ni} of pairs involving that group are recalculated, and Eqs.~\ref{eq:total} and \ref{eq:Ci} are updated accordingly.

To compare the global dust evolution using the two algorithms, we ran independent simulations with the same setup and resolution. We initialized a disk of $100\,\mathrm{AU}$ in size with a static gas surface density and temperature given by the power laws 
\begin{eqnarray}
\label{eq:sigma}
\Sigma_{g} &=& 800\left( \frac{r}{\mathrm{AU}}\right)^{-1}\,\mathrm{g\,cm^{-2}}\,, \\
T&=& 280\left( \frac{r}{\mathrm{AU}}\right)^{-1/2}\,\mathrm{K}\,,
\end{eqnarray}
where $r$ is the radial distance from a Sun-like star. The turbulence parameter was set to $\alpha\,$$=\,$$10^{-3}$. For the vertical structure, we assumed a Gaussian gas density profile and an isothermal temperature structure. We employed $256$ representative particles in each bin in the collision grid, with a total of $200$ radial and $20$ vertical bins. We began the simulation with a global vertically integrated dust-to-gas ratio of $0.01$, with monomers of $1\,\mathrm{\mu m}$ in radius and an internal density of $1\,\mathrm{g\,cm^{-3}}$. We assumed that particles fragment when they collide at velocities above $10\,\mathrm{m\,s^{-1}}$. Studies on water-ice aggregates support this threshold value \citep[e.g.,][]{Wada2013, Gundlach2015}, although its temperature dependency remains under debate \citep[e.g.,][]{Gundlach2018, Musiolik2019}. While silicates have a higher internal density than water ice and are expected to fragment at lower velocities \citep{Guttler2010}, we simplified our approach by assuming a single component throughout the disk.

Figure \ref{fig:mcdust} compares the two methods at a snapshot of $10^{4}\,\mathrm{yr}$. At this point, particles in the inner regions of the disk had already grown to sizes at which fragmentation occurs. In the outer regions, where collisional timescales are longer, dust had not yet reached any growth barrier. Since particles initially grow by adding one monomer at a time, there are no particles with sizes between $1$ and $2\,\mathrm{\mu m}$ in the outer regions (beyond $20-30\,\mathrm{AU}$). In contrast, in the inner regions fragmentation replenishes the population of $1$–$2\,\mathrm{\mu m}$ particles, because, as previously stated, fragmentation is not limited to discrete monomers in this work.

Overall, the two methods yield very similar results, and many of the small differences between the snapshots in Fig.~\ref{fig:mcdust} are attributable to the stochastic nature of the Monte Carlo simulations. The new method achieves modest improvement in the resolution of small particles. In the approach of \cite{Zsom2008}, each representative particle carries the same total mass, which results in small grains being represented by only a few representative particles. In contrast, our method allows the total mass of each group to vary, enabling a larger number of representative particles to sample the small-size regime. The trade-off is a reduced number of representative particles at the largest sizes. The merging parameter $X$ (see Sect.~\ref{sec:filling}) plays a central role in this balance: low values of $X$ enable larger variation in group mass, resulting in fewer merging events and consequently splitting large groups less often. For applications such as the synthetic observations of protoplanetary disks, where the opacity may be dominated by small dust grains, lower values of $X$ are therefore particularly suitable, although $X\,$$<\,$$0.01$ requires a particle resolution of more than $200$ particles per grid cell (see Sect.~\ref{sec:kernel}). In this respect, the method presented here enables greater flexibility. 

All simulations in Fig.~\ref{fig:mcdust} were evolved up to $10^{4}\,\mathrm{yr}$ on identical computational resources (128 CPU cores), each requiring approximately $3\,\mathrm{h}$ of wall-clock time. For the setups considered in this study, the two algorithms therefore exhibit comparable computational cost.

\begin{figure*}
    \centering
    \includegraphics[width=17cm]{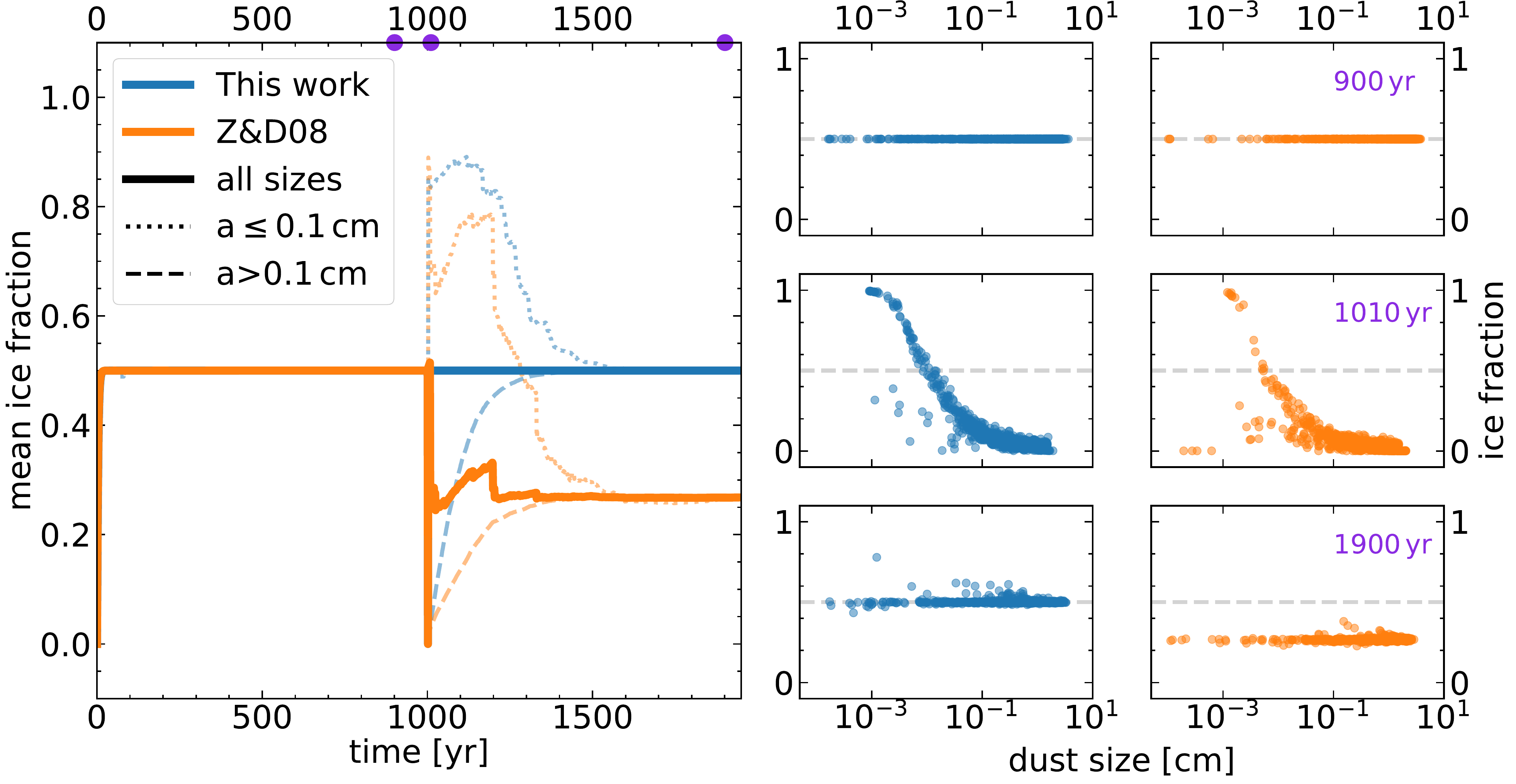}
    \caption{Left: Mean water ice fraction of the dust population before and after an outburst-like event triggered at $1000\,\mathrm{yr}$. Two independent simulations are shown: the coagulation algorithm (blue line) presented in this work and the algorithm of \citet{Zsom2008} (orange line). The solid line indicates the average ice fraction over all particle sizes, the dotted line the average for small particles, and the dashed line the average for large particles. Initially, water vapor condenses almost instantaneously into particles of equal size. Later, ice sublimates during the outburst event that lasts $1\,\mathrm{yr}$. Upon recondensation onto silicate grains of varying sizes, the coagulation method presented here exactly conserves the mean ice fraction, while the method by \citet{Zsom2008} does not. Right: Water ice fraction and sizes of every representative particle at different times for the two simulations. Before the outburst ($900\,\mathrm{yr}$), water is equally distributed across particles of different sizes. Following the outburst ($1010\,\mathrm{yr}$), small particles gain more water in proportion. At later times ($1900\,\mathrm{yr}$), the water fraction is again nearly equally distributed across different sizes.}
     \label{fig:mass}
\end{figure*}

\subsection{Conservation of the global budget: water and silicates}\label{sec:water}

To demonstrate that our algorithm conserves the global budget of particle properties, we built the following toy model. We considered a zero-dimensional simulation with $10^{3}$ representative particles composed of silicates and water ice at a radial distance of $5\,\mathrm{AU}$. For the same disk parameters as in Sect.~\ref{sec:global}, this location lies beyond the water ice line. We triggered an outburst-like event by increasing the disk temperature by a factor of $3$. The increase in temperature shifts the water-ice line beyond $5\,\mathrm{AU}$, leading to ice sublimation in the simulated region. We assumed that the silicate aggregates survive sublimation without fragmenting back into monomers \citep{Houge2023, Houge2024}. After the outburst, the water vapor recondenses back onto the silicate grains.

The recondensation of water onto dust grains within a volume $V$ occurs when the vapor pressure in that volume exceeds the saturation vapor pressure. The volume density for saturation is \citep{Supulver2000},
\begin{equation}\label{eq:saturation}
    \rho_{\rm sat} = \frac{m_{H_2O}}{k_{B}T} \cdot  1.013\times 10^{6}\exp{\left[ 15.6-\frac{5940\,\mathrm{K}}{T}\right]}\,\mathrm{dyn/cm^{2}}\,,
\end{equation}
where $m_{H_2O}$ is the mass of water molecules. Vapor deposition onto an existing icy surface is energetically more favorable than the formation of an initial ice monolayer on a silicate grain \citep{Ros2019}. Although the distinction between icy and non-icy particles during recondensation can have strong implications on the dust size distribution after an outburst \citep{Ros2024}, for this test case, we assumed that water condensation is independent of the grain's surface composition. We included size-dependent condensation and evaporation of water on dust particles, following the approach of \citet{Krijt2016}. Let $\rho_{H_2O}$ be the volume density of water vapor. When $\rho_{H_2O}\,$$>\,$$\rho_{\rm sat}$, every particle gains water ice, $m_{\mathrm{ice}, i}$, over the global time step $\Delta t_{\rm global}$,
\begin{equation} \label{eq:condensation}
\Delta m_{\mathrm{ice}, i} =(1-e^{\Delta t_{\rm global}/\tau_{\mathrm{con}}})\left( \rho_{H_{2}O}-\rho_{\rm sat}\right) \frac{a^{2}_{i}}{\sum_{i=1}^{N_{r}} N_{i} a^{2}_{i}} V\,,
\end{equation}
where $\tau_{\rm con}$ is the condensation timescale, $a_{i}$ is the particle radius, and $N_{i}$ is the number of physical particles in group $\{i\}$. $\tau_{\rm con}$ is defined as 
\begin{equation}\label{eq:timescale_cond}
    \tau_{\rm con} = \frac{V}{ \pi v_{\mathrm{th}}\sum_{i=1}^{N_{r}} N_{i}a^{2}_{i} }\,,
\end{equation}
where $v_{\rm th}=(8k_{B}T/\pi m_{H_2O})^{1/2}$ is the thermal velocity of water molecules. When $\rho_{H_2O}\,$$<\,$$\rho_{\rm sat}$, every particle loses water ice over the time step $\Delta t_{\rm global}$ as
\begin{equation} \label{eq:sublimation}
    \Delta m_{\mathrm{ice}, i} = - \min\left(\pi a^{2}_{i}v_{th}\left( \rho_{\rm sat}-\rho_{H_2O}\right)\Delta t_{\rm global},  \;m_{\mathrm{ice}, i}\right)\,.
\end{equation}
Consequently, the total particle mass and radius, $m_i$ and $a_i$, increase or decrease during condensation or sublimation. Since in the method by \cite{Zsom2008} the total mass of every representative particle, $M_{i}\,$$=\,$$N_{i}m_{i}$, is equal, if mass changes, the number of physical particles $N_{i}$ must change accordingly. However, during sublimation and recondensation the particle number should remain constant. Therefore, a more appropriate assumption is to keep the total silicate mass of every particle, $M_{ \mathrm{si},i}\,$$=\,$$N_{i}m_{ \mathrm{si},i}$, constant, since $m_{ \mathrm{si},i}$ remains constant during condensation or sublimation. In that case, the total silicate mass is conserved, whereas the total mass of water may fluctuate during collisions. In addition, maintaining $M_{ \mathrm{si},i}$ constant does not allow for tracking particles that contain only ice. On the other hand, the method presented in this work does not have this limitation, as $M_{i}$ is allowed to vary.

To test the distribution of water during collisions after an outburst, we ran two independent simulations: one using the coagulation algorithm presented in this work, and another using the algorithm of \citet{Zsom2008}. We considered sticking and fragmentation during collisions, as described in Sect.~\ref{sec:global}, with the outcome determined by the fragmentation velocity of $10\,\mathrm{m\,s^{-1}}$. In reality, silicate aggregates may fragment at velocities an order of magnitude lower \citep{Guttler2010}. However, we adopted a single fragmentation threshold to simplify the model.

Throughout the simulations, we tracked the total particle mass, $m_{i}$, along with the ice fraction of dust particles, defined as
\begin{equation}
    f_{i}= \frac{M_{\mathrm{ice},i}}{M_{i}} = \frac{m_{\mathrm{ice},i}}{m_{i}}\,.
\end{equation}
We then calculated the component masses as $m_{\mathrm{ice},i}\,$$=\,$$f_{i}m_{i}$ and $m_{\mathrm{si},i}\,$$=\,$$(1-f_{i})\,m_i$. In the algorithm presented in this paper, during sticking collisions between group $\{i\}$ and $\{j\}$, we update the properties as
\begin{eqnarray}
\label{eq:ice_fraction}
f^{\prime}_{i}&=& \frac{M^{\prime}_{\mathrm{ice},i}}{M^{\prime}_{i}}=\frac{N_{i}m_{i}f_{i}+N_{i}m_{j}f_{j}}{N_{i}m_{i}+N_{i}m_{j}}= \frac{m_{i}f_{i}+m_{j}f_{j}}{m_{i}+m_{j}}\,, \\
f^{\prime}_{j}&=& \frac{M^{\prime}_{\mathrm{ice},j}}{M^{\prime}_{j}}=\frac{N_{j}m_{j}f_{j}-N_{i}m_{j}f_{j}}{N_{j}m_{j}-N_{i}m_{j}}=  f_{j}\,,
\end{eqnarray}
and thus the total ice mass budget $M_{\rm{ice},i}+M_{\rm{ice},j}$ in a sticking event is conserved (see Eqs.~\ref{eq:stick}-\ref{eq:last}). If group $\{j\}$ contains negligible mass after transferring mass to group $\{i\}$ (see Eq.~\ref{eq:merging}), it merges with its most similar group $\{w\}$ (see Eq.~\ref{eq:similar_water}), and the resulting ice fraction is
\begin{equation}
f^{\prime}_{w}= \frac{M^{\prime}_{\mathrm{ice},w}}{M^{\prime}_{w}}
= \frac{N_{w} m_{w} f_{w} + N_{j} m_{j} f_{j}}
{N_{w} m_{w} + N_{j} m_{j}}.
\end{equation}
We chose a merging parameter of $X\,$$=\,$$0.01$ for this simulation.

Assuming that the aggregates are compact, spherical, and well mixed in composition, the internal density depends on the ice fraction as well as the ice and silicate densities as
\begin{equation}
    \rho_{i} = \left( \frac{f_{i}}{\rho_{\rm ice}} + \frac{1-f_{i}}{\rho_{\rm silicate}}\right)^{-1}\,.
\end{equation}
We set $\rho_{\rm ice}=1\,\mathrm{g\,cm^{-3}}$ and $\rho_{\rm silicate}=3\,\mathrm{g\,cm^{-3}}$. In reality, aggregates may not be well mixed, and consequently, the composition of fragments can vary \citep{Krijt2024}. Although Monte Carlo methods can track variations in fragment composition \citep[e.g.,][]{Gurrutxaga2026}, further refinement is required to ensure conservation of each component's total mass to track heterogeneous fragmentation. In this study, we neglected this and assumed that the ice fraction, $f_{i}$, remains constant during fragmentation.

In this setup, particles are expected to coagulate to centimeter sizes before fragmenting. We therefore initialized the simulation with centimeter-sized silicate particles settled in the midplane, corresponding to an initial Stokes number of $\rm{St}\,$$\approx\,$$0.03$. For a vertically integrated dust-to-gas ratio of $Z_{\rm si}\,$$=\,$$0.005$ and turbulence of $\alpha\,$$=\,$$10^{-3}$, the midplane dust-to-gas ratio of these dust population after vertical settling is \citep[][]{Youdin2007}
\begin{equation}\label{eq:Zmidplane}
\epsilon_{\rm si} = \sqrt{\frac{\alpha + \rm{St}}{\rm{\alpha}}} Z_{\rm{si}} \approx 0.028\,,
\end{equation}
which was expected to remain roughly constant for our simulated test case. We thus assumed a constant volume $V$ and a fixed number of representative particles. We also assumed that the initial vapor density at the midplane equals the dust volume density.

After initializing the simulation, we evolved the dust population for $1000\,\mathrm{yr}$ to reach a coagulation–fragmentation equilibrium. We then triggered a brief outburst of $1\,\mathrm{yr}$ and let the simulation stabilize again for another $1000\,\mathrm{yr}$. We neglected dust transport, and therefore the global time step $\Delta t_{\rm global}$ was set solely by the collisional timescale (see Sect.~\ref{sec:global}). However, we imposed an additional constraint on the global timescale by setting a maximum of $1\,\mathrm{yr}$ to resolve the temperature change due to the outburst and the resulting sublimation and recondensation.

Figure \ref{fig:mass} shows that for our disk parameters at $5\,\mathrm{AU}$, water condenses almost instantaneously onto centimeter-sized silicate grains, since our initial condensation timescale is $\sim\,$$7\,\mathrm{yr}$ (see Eq.~\ref{eq:timescale_cond}). Because all particles initially have the same size, each silicate grain gains the same water content, and the total ice mass is conserved in both algorithms during coagulation (see Eq.~\ref{eq:condensation}). Subsequently, an outburst-like event is triggered at $1000\,\mathrm{yr}$. The rise in temperature increases the saturation vapor density (see Eq.~\ref{eq:saturation}), and consequently, water ice sublimates. After the outburst, water vapor recondenses onto the silicate grains. At this stage, particle sizes are no longer equal, and smaller grains acquire a higher ice mass fraction (see Eq.~\ref{eq:condensation}; and also Fig.~\ref{fig:mass}, right panels). In both simulations considered here, the dust population requires approximately $500\,\mathrm{yr}$ to redistribute the ice uniformly after the outburst (see Fig.~\ref{fig:mass}, dashed and dotted lines), which is faster than the time reported by \cite{Houge2023} for their resilient model. This difference in re-equilibration time is likely due to our choice of a higher midplane dust-to-gas ratio (see Eq.~\ref{eq:Zmidplane}).

During collisions between particles with different ice contents, the algorithm of \cite{Zsom2008} updates only one of the colliding representative particles and therefore does not conserve the total water content, leading to numerical fluctuations (see Fig.~\ref{fig:mass}, solid orange line). While these fluctuations would decrease with increasing particle number \citep{Houge2023}, simulations with such high numbers of representative particles per grid cell in global disks are currently computationally too expensive. By contrast, we demonstrate that the new algorithm conserves water and silicate mass independently, without introducing such fluctuations.

\section{Conclusions}\label{sec:conclusion}
We presented a Monte Carlo method for computing dust coagulation in protoplanetary disks. Although derived independently, the method is conceptually equivalent to the Monte Carlo approach introduced by \cite{Shima2009} for cloud microphysics. A key strength of this approach is its ability to conserve the global inventory of dust properties while remaining applicable to global disk simulations.

We first validated the algorithm by benchmarking it against analytical solutions of the Smoluchowski equation. We demonstrated that the method successfully reproduces the analytical results and that it requires a minimum of $200$ representative particles (see Figs.~\ref{fig:kernel} and \ref{fig:kernel_lowres}).

We then implemented the method in a two-dimensional protoplanetary disk model and compared it with the algorithm of \citet{Zsom2008}. We find that both approaches yield similar results when the same numerical resolution is employed (see Fig.~\ref{fig:mcdust}). We introduced the merging parameter $X$ in Sect.~\ref{sec:filling}, which can be adjusted to concentrate resolution on the largest particles (e.g., $X\,$$\geq\,$$0.1$) or to distribute it more evenly across the size distribution (e.g., $X\,$$<\,$$0.1$). We note, however, that decreasing $X$ requires a larger number of representative particles to resolve dust evolution, since collisions are typically dominated by the largest particles in a protoplanetary disk.

By tracking the composition of a dust population undergoing sublimation and recondensation events, we demonstrated that the method presented here conserves the global budget of particle properties (see Fig.~\ref{fig:mass}). Overall, this method offers a promising tool for tracking dust properties during coagulation in global protoplanetary disk simulations.

\begin{acknowledgements}
The authors thank the anonymous referee for constructive comments that helped to improve the manuscript. J.D. and V.V. are funded by the European Union under the European Union’s Horizon Europe Research \& Innovation Programme 101040037 (PLANETOIDS).
\end{acknowledgements}
\section*{Data and code availability}
All data and code required to reproduce the results presented in this work are publicly available. The dataset can be accessed via Zenodo at \href{https://zenodo.org/records/19390275}{https://doi.org/10.5281/zenodo.19390275}. The code used to perform the simulations and generate all figures is available in the GitHub repository at \href{https://github.com/nereagurru/A-Monte-Carlo-method-for-tracking-dust-properties-during-coagulation-in-protoplanetary-disks}{https://github.com/nereagurru/A-Monte-Carlo-method-for-tracking-dust-properties-during-coagulation-in-protoplanetary-disks}. The original \texttt{mcdust} codebase is available at \href{https://github.com/vicky1997/mcdust}{https://github.com/vicky1997/mcdust}.

\bibliographystyle{bibtex/aa}
\bibliography{bibtex/biblio}

@ARTICLE{Drazkowska2013,
       author = {{Dr{\k{a}}{\.z}kowska}, J. and {Windmark}, F. and {Dullemond}, C.~P.},
        title = "{Planetesimal formation via sweep-up growth at the inner edge of dead zones}",
      journal = {\aap},
         year = 2013,
        month = aug,
       volume = {556},
          eid = {A37},
        pages = {A37},
          doi = {10.1051/0004-6361/201321566},
archivePrefix = {arXiv},
       eprint = {1306.3412},
 primaryClass = {astro-ph.EP},
       adsurl = {https://ui.adsabs.harvard.edu/abs/2013A&A...556A..37D}
}

@ARTICLE{Vaikundaraman2025,
       author = {{Vaikundaraman}, Vignesh and {Gurrutxaga}, Nerea and {Dr{\k{a}}{\.z}kowska}, Joanna},
        title = "{mcdust: A 2D Monte Carlo code for dust coagulation in protoplanetary disks}",
      journal = {arXiv e-prints},
         year = 2025,
        month = jul,
          eid = {arXiv:2507.21239},
        pages = {arXiv:2507.21239},
          doi = {10.48550/arXiv.2507.21239},
archivePrefix = {arXiv},
       eprint = {2507.21239},
 primaryClass = {astro-ph.IM},
       adsurl = {https://ui.adsabs.harvard.edu/abs/2025arXiv250721239V}
}

@ARTICLE{Vaikundaraman2025_master,
       author = {{Vaikundaraman}, Vignesh and {Dr{\k{a}}{\.z}kowska}, Joanna and {Binkert}, Fabian and {Birnstiel}, Til and {Miotello}, Anna},
        title = "{Refractory carbon depletion by photolysis through dust collisions and vertical mixing}",
      journal = {\aap},
         year = 2025,
        month = apr,
       volume = {696},
          eid = {A215},
        pages = {A215},
          doi = {10.1051/0004-6361/202553850},
archivePrefix = {arXiv},
       eprint = {2503.14597},
 primaryClass = {astro-ph.EP},
       adsurl = {https://ui.adsabs.harvard.edu/abs/2025A&A...696A.215V}
}

@ARTICLE{Krijt2016,
       author = {{Krijt}, Sebastiaan and {Ciesla}, Fred J. and {Bergin}, Edwin A.},
        title = "{Tracing Water Vapor and Ice During Dust Growth}",
      journal = {\apj},
         year = 2016,
        month = dec,
       volume = {833},
       number = {2},
          eid = {285},
        pages = {285},
          doi = {10.3847/1538-4357/833/2/285},
archivePrefix = {arXiv},
       eprint = {1610.06463},
 primaryClass = {astro-ph.EP},
       adsurl = {https://ui.adsabs.harvard.edu/abs/2016ApJ...833..285K}
}

@ARTICLE{Houge2023,
       author = {{Houge}, Adrien and {Krijt}, Sebastiaan},
        title = "{Collisional evolution of dust and water ice in protoplanetary discs during and after an accretion outburst}",
      journal = {\mnras},
         year = 2023,
        month = jun,
       volume = {521},
       number = {4},
        pages = {5826-5845},
          doi = {10.1093/mnras/stad866},
archivePrefix = {arXiv},
       eprint = {2303.11318},
 primaryClass = {astro-ph.EP},
       adsurl = {https://ui.adsabs.harvard.edu/abs/2023MNRAS.521.5826H}
}

@ARTICLE{Ormel2007,
       author = {{Ormel}, C.~W. and {Spaans}, M. and {Tielens}, A.~G.~G.~M.},
        title = "{Dust coagulation in protoplanetary disks: porosity matters}",
      journal = {\aap},
         year = 2007,
        month = jan,
       volume = {461},
       number = {1},
        pages = {215-232},
          doi = {10.1051/0004-6361:20065949},
archivePrefix = {arXiv},
       eprint = {astro-ph/0610030},
 primaryClass = {astro-ph},
       adsurl = {https://ui.adsabs.harvard.edu/abs/2007A&A...461..215O}
}

@ARTICLE{Ormel2008,
       author = {{Ormel}, C.~W. and {Spaans}, M.},
        title = "{Monte Carlo Simulation of Particle Interactions at High Dynamic Range: Advancing beyond the Googol}",
      journal = {\apj},
         year = 2008,
        month = sep,
       volume = {684},
       number = {2},
        pages = {1291-1309},
          doi = {10.1086/590052},
archivePrefix = {arXiv},
       eprint = {0804.4449},
 primaryClass = {astro-ph},
       adsurl = {https://ui.adsabs.harvard.edu/abs/2008ApJ...684.1291O}
}

@ARTICLE{Zsom2008,
       author = {{Zsom}, A. and {Dullemond}, C.~P.},
        title = "{A representative particle approach to coagulation and fragmentation of dust aggregates and fluid droplets}",
      journal = {\aap},
         year = 2008,
        month = oct,
       volume = {489},
       number = {2},
        pages = {931-941},
          doi = {10.1051/0004-6361:200809921},
archivePrefix = {arXiv},
       eprint = {0807.5052},
 primaryClass = {astro-ph},
       adsurl = {https://ui.adsabs.harvard.edu/abs/2008A&A...489..931Z}
}

@ARTICLE{Tanaka1994,
       author = {{Tanaka}, H. and {Nakazawa}, K.},
        title = "{Validity of the Statistical Coagulation Equation and Runaway Growth of Protoplanets}",
      journal = {\icarus},
         year = 1994,
        month = feb,
       volume = {107},
       number = {2},
        pages = {404-412},
          doi = {10.1006/icar.1994.1032},
       adsurl = {https://ui.adsabs.harvard.edu/abs/1994Icar..107..404T}
}

@ARTICLE{Liffman1992,
       author = {{Liffman}, Kurt},
        title = "{A Direct Simulation Monte-Carlo Method for Cluster Coagulation}",
      journal = {Journal of Computational Physics},
         year = 1992,
        month = may,
       volume = {100},
       number = {1},
        pages = {116-127},
          doi = {10.1016/0021-9991(92)90314-O},
       adsurl = {https://ui.adsabs.harvard.edu/abs/1992JCoPh.100..116L}
}

@ARTICLE{Gillespie1975,
       author = {{Gillespie}, Daniel T.},
        title = "{An Exact Method for Numerically Simulating the Stochastic Coalescence Process in a Cloud.}",
      journal = {Journal of the Atmospheric Sciences},
         year = 1975,
        month = oct,
       volume = {32},
       number = {10},
        pages = {1977-1989},
          doi = {10.1175/1520-0469(1975)032<1977:AEMFNS>2.0.CO;2},
       adsurl = {https://ui.adsabs.harvard.edu/abs/1975JAtS...32.1977G}
}

@ARTICLE{Schrapler2011,
       author = {{Schr{\"a}pler}, Rainer and {Blum}, J{\"u}rgen},
        title = "{The Physics of Protoplanetesimal Dust Agglomerates. VI. Erosion of Large Aggregates as a Source of Micrometer-sized Particles}",
      journal = {\apj},
         year = 2011,
        month = jun,
       volume = {734},
       number = {2},
          eid = {108},
        pages = {108},
          doi = {10.1088/0004-637X/734/2/108},
archivePrefix = {arXiv},
       eprint = {1103.0427},
 primaryClass = {astro-ph.EP},
       adsurl = {https://ui.adsabs.harvard.edu/abs/2011ApJ...734..108S}
}

@ARTICLE{Beutel2023,
       author = {{Beutel}, M. and {Dullemond}, C.~P.},
        title = "{An improved Representative Particle Monte Carlo method for the simulation of particle growth}",
      journal = {\aap},
         year = 2023,
        month = feb,
       volume = {670},
          eid = {A134},
        pages = {A134},
          doi = {10.1051/0004-6361/202244955},
       adsurl = {https://ui.adsabs.harvard.edu/abs/2023A&A...670A.134B}
}

@ARTICLE{Weidenschilling1980,
       author = {{Weidenschilling}, S.~J.},
        title = "{Dust to planetesimals: Settling and coagulation in the solar nebula}",
      journal = {\icarus},
         year = 1980,
        month = oct,
       volume = {44},
       number = {1},
        pages = {172-189},
          doi = {10.1016/0019-1035(80)90064-0},
       adsurl = {https://ui.adsabs.harvard.edu/abs/1980Icar...44..172W}
}

@ARTICLE{Dullemond2005,
       author = {{Dullemond}, C.~P. and {Dominik}, C.},
        title = "{Dust coagulation in protoplanetary disks: A rapid depletion of small grains}",
      journal = {\aap},
         year = 2005,
        month = may,
       volume = {434},
       number = {3},
        pages = {971-986},
          doi = {10.1051/0004-6361:20042080},
archivePrefix = {arXiv},
       eprint = {astro-ph/0412117},
 primaryClass = {astro-ph},
       adsurl = {https://ui.adsabs.harvard.edu/abs/2005A&A...434..971D}
}

@ARTICLE{Brauer2008,
       author = {{Brauer}, F. and {Dullemond}, C.~P. and {Henning}, Th.},
        title = "{Coagulation, fragmentation and radial motion of solid particles in protoplanetary disks}",
      journal = {\aap},
         year = 2008,
        month = mar,
       volume = {480},
       number = {3},
        pages = {859-877},
          doi = {10.1051/0004-6361:20077759},
archivePrefix = {arXiv},
       eprint = {0711.2192},
 primaryClass = {astro-ph},
       adsurl = {https://ui.adsabs.harvard.edu/abs/2008A&A...480..859B}
}

@ARTICLE{Birnstiel2010,
       author = {{Birnstiel}, T. and {Dullemond}, C.~P. and {Brauer}, F.},
        title = "{Gas- and dust evolution in protoplanetary disks}",
      journal = {\aap},
         year = 2010,
        month = apr,
       volume = {513},
          eid = {A79},
        pages = {A79},
          doi = {10.1051/0004-6361/200913731},
archivePrefix = {arXiv},
       eprint = {1002.0335},
 primaryClass = {astro-ph.EP},
       adsurl = {https://ui.adsabs.harvard.edu/abs/2010A&A...513A..79B}
}

@ARTICLE{Stammler2022,
       author = {{Stammler}, Sebastian M. and {Birnstiel}, Tilman},
        title = "{DustPy: A Python Package for Dust Evolution in Protoplanetary Disks}",
      journal = {\apj},
         year = 2022,
        month = aug,
       volume = {935},
       number = {1},
          eid = {35},
        pages = {35},
          doi = {10.3847/1538-4357/ac7d58},
archivePrefix = {arXiv},
       eprint = {2207.00322},
 primaryClass = {astro-ph.EP},
       adsurl = {https://ui.adsabs.harvard.edu/abs/2022ApJ...935...35S}
}

@ARTICLE{Ossenkopf1993,
       author = {{Ossenkopf}, V.},
        title = "{Dust coagulation in dense molecular clouds : the formation of fluffy aggregates.}",
      journal = {\aap},
         year = 1993,
        month = dec,
       volume = {280},
        pages = {617-646},
       adsurl = {https://ui.adsabs.harvard.edu/abs/1993A&A...280..617O}
}

@ARTICLE{Okuzumi2009,
       author = {{Okuzumi}, Satoshi and {Tanaka}, Hidekazu and {Sakagami}, Masa-aki},
        title = "{Numerical Modeling of the Coagulation and Porosity Evolution of Dust Aggregates}",
      journal = {\apj},
         year = 2009,
        month = dec,
       volume = {707},
       number = {2},
        pages = {1247-1263},
          doi = {10.1088/0004-637X/707/2/1247},
archivePrefix = {arXiv},
       eprint = {0911.0239},
 primaryClass = {astro-ph.EP},
       adsurl = {https://ui.adsabs.harvard.edu/abs/2009ApJ...707.1247O}
}

@ARTICLE{Stammler2017,
       author = {{Stammler}, Sebastian Markus and {Birnstiel}, Tilman and {Pani{\'c}}, Olja and {Dullemond}, Cornelis Petrus and {Dominik}, Carsten},
        title = "{Redistribution of CO at the location of the CO ice line in evolving gas and dust disks}",
      journal = {\aap},
         year = 2017,
        month = apr,
       volume = {600},
          eid = {A140},
        pages = {A140},
          doi = {10.1051/0004-6361/201629041},
archivePrefix = {arXiv},
       eprint = {1701.02385},
 primaryClass = {astro-ph.EP},
       adsurl = {https://ui.adsabs.harvard.edu/abs/2017A&A...600A.140S}
}

@ARTICLE{Youdin2005,
       author = {{Youdin}, Andrew N. and {Goodman}, Jeremy},
        title = "{Streaming Instabilities in Protoplanetary Disks}",
      journal = {\apj},
         year = 2005,
        month = feb,
       volume = {620},
       number = {1},
        pages = {459-469},
          doi = {10.1086/426895},
archivePrefix = {arXiv},
       eprint = {astro-ph/0409263},
 primaryClass = {astro-ph},
       adsurl = {https://ui.adsabs.harvard.edu/abs/2005ApJ...620..459Y}
}

@ARTICLE{Ormel2010,
       author = {{Ormel}, C.~W. and {Klahr}, H.~H.},
        title = "{The effect of gas drag on the growth of protoplanets. Analytical expressions for the accretion of small bodies in laminar disks}",
      journal = {\aap},
         year = 2010,
        month = sep,
       volume = {520},
          eid = {A43},
        pages = {A43},
          doi = {10.1051/0004-6361/201014903},
archivePrefix = {arXiv},
       eprint = {1007.0916},
 primaryClass = {astro-ph.EP},
       adsurl = {https://ui.adsabs.harvard.edu/abs/2010A&A...520A..43O}
}

@ARTICLE{Lambrechts2012,
       author = {{Lambrechts}, M. and {Johansen}, A.},
        title = "{Rapid growth of gas-giant cores by pebble accretion}",
      journal = {\aap},
         year = 2012,
        month = aug,
       volume = {544},
          eid = {A32},
        pages = {A32},
          doi = {10.1051/0004-6361/201219127},
archivePrefix = {arXiv},
       eprint = {1205.3030},
 primaryClass = {astro-ph.EP},
       adsurl = {https://ui.adsabs.harvard.edu/abs/2012A&A...544A..32L}
}

@ARTICLE{Birnstiel2024,
       author = {{Birnstiel}, Tilman},
        title = "{Dust Growth and Evolution in Protoplanetary Disks}",
      journal = {\araa},
         year = 2024,
        month = sep,
       volume = {62},
       number = {1},
        pages = {157-202},
          doi = {10.1146/annurev-astro-071221-052705},
archivePrefix = {arXiv},
       eprint = {2312.13287},
 primaryClass = {astro-ph.EP},
       adsurl = {https://ui.adsabs.harvard.edu/abs/2024ARA&A..62..157B}
}

@ARTICLE{Robinson2024,
       author = {{Robinson}, Alfie and {Booth}, Richard A. and {Owen}, James E.},
        title = "{Introducing CUDISC: a 2D code for protoplanetary disc structure and evolution calculations}",
      journal = {\mnras},
         year = 2024,
        month = apr,
       volume = {529},
       number = {2},
        pages = {1524-1541},
          doi = {10.1093/mnras/stae624},
archivePrefix = {arXiv},
       eprint = {2402.18471},
 primaryClass = {astro-ph.EP},
       adsurl = {https://ui.adsabs.harvard.edu/abs/2024MNRAS.529.1524R}
}

@ARTICLE{Birnstiel2012,
       author = {{Birnstiel}, T. and {Klahr}, H. and {Ercolano}, B.},
        title = "{A simple model for the evolution of the dust population in protoplanetary disks}",
      journal = {\aap},
         year = 2012,
        month = mar,
       volume = {539},
          eid = {A148},
        pages = {A148},
          doi = {10.1051/0004-6361/201118136},
archivePrefix = {arXiv},
       eprint = {1201.5781},
 primaryClass = {astro-ph.EP},
       adsurl = {https://ui.adsabs.harvard.edu/abs/2012A&A...539A.148B}
}

@ARTICLE{Akimkin2023,
       author = {{Akimkin}, Vitaly and {Ivlev}, Alexei V. and {Caselli}, Paola and {Gong}, Munan and {Silsbee}, Kedron},
        title = "{Coagulation-Fragmentation Equilibrium for Charged Dust: Abundance of Submicron Grains Increases Dramatically in Protoplanetary Disks}",
      journal = {\apj},
         year = 2023,
        month = aug,
       volume = {953},
       number = {1},
          eid = {72},
        pages = {72},
          doi = {10.3847/1538-4357/ace2c5},
archivePrefix = {arXiv},
       eprint = {2306.16408},
 primaryClass = {astro-ph.SR},
       adsurl = {https://ui.adsabs.harvard.edu/abs/2023ApJ...953...72A}
}

@ARTICLE{Guttler2010,
       author = {{G{\"u}ttler}, C. and {Blum}, J. and {Zsom}, A. and {Ormel}, C.~W. and {Dullemond}, C.~P.},
        title = "{The outcome of protoplanetary dust growth: pebbles, boulders, or planetesimals?. I. Mapping the zoo of laboratory collision experiments}",
      journal = {\aap},
         year = 2010,
        month = apr,
       volume = {513},
          eid = {A56},
        pages = {A56},
          doi = {10.1051/0004-6361/200912852},
archivePrefix = {arXiv},
       eprint = {0910.4251},
 primaryClass = {astro-ph.EP},
       adsurl = {https://ui.adsabs.harvard.edu/abs/2010A&A...513A..56G}
}

@ARTICLE{Blum1993,
       author = {{Blum}, J{\"u}rgen and {M{\"u}nch}, Michael},
        title = "{Experimental Investigations on Aggregate-Aggregate Collisions in the Early Solar Nebula}",
      journal = {\icarus},
         year = 1993,
        month = nov,
       volume = {106},
       number = {1},
        pages = {151-167},
          doi = {10.1006/icar.1993.1163},
       adsurl = {https://ui.adsabs.harvard.edu/abs/1993Icar..106..151B}
}

@ARTICLE{Zsom2010,
       author = {{Zsom}, A. and {Ormel}, C.~W. and {G{\"u}ttler}, C. and {Blum}, J. and {Dullemond}, C.~P.},
        title = "{The outcome of protoplanetary dust growth: pebbles, boulders, or planetesimals? II. Introducing the bouncing barrier}",
      journal = {\aap},
         year = 2010,
        month = apr,
       volume = {513},
          eid = {A57},
        pages = {A57},
          doi = {10.1051/0004-6361/200912976},
archivePrefix = {arXiv},
       eprint = {1001.0488},
 primaryClass = {astro-ph.EP},
       adsurl = {https://ui.adsabs.harvard.edu/abs/2010A&A...513A..57Z}
}

@ARTICLE{Kelling2014,
       author = {{Kelling}, T. and {Wurm}, G. and {K{\"o}ster}, M.},
        title = "{Experimental Study on Bouncing Barriers in Protoplanetary Disks}",
      journal = {\apj},
         year = 2014,
        month = mar,
       volume = {783},
       number = {2},
          eid = {111},
        pages = {111},
          doi = {10.1088/0004-637X/783/2/111},
archivePrefix = {arXiv},
       eprint = {1401.4280},
 primaryClass = {astro-ph.EP},
       adsurl = {https://ui.adsabs.harvard.edu/abs/2014ApJ...783..111K}
}

@ARTICLE{Smoluchowski1916,
       author = {{Smoluchowski}, M.~V.},
        title = "{Drei Vortrage uber Diffusion, Brownsche Bewegung und Koagulation von Kolloidteilchen}",
      journal = {Zeitschrift fur Physik},
         year = 1916,
        month = jan,
       volume = {17},
        pages = {557-585},
       adsurl = {https://ui.adsabs.harvard.edu/abs/1916ZPhy...17..557S}
}

@ARTICLE{Houge2024,
       author = {{Houge}, Adrien and {Mac{\'\i}as}, Enrique and {Krijt}, Sebastiaan},
        title = "{Surviving the heat: multiwavelength analysis of V883 Ori reveals that dust aggregates survive the sublimation of their ice mantles}",
      journal = {\mnras},
         year = 2024,
        month = feb,
       volume = {527},
       number = {4},
        pages = {9668-9682},
          doi = {10.1093/mnras/stad3758},
archivePrefix = {arXiv},
       eprint = {2312.01856},
 primaryClass = {astro-ph.EP},
       adsurl = {https://ui.adsabs.harvard.edu/abs/2024MNRAS.527.9668H}
}

@ARTICLE{Krijt2015,
       author = {{Krijt}, S. and {Ormel}, C.~W. and {Dominik}, C. and {Tielens}, A.~G.~G.~M.},
        title = "{Erosion and the limits to planetesimal growth}",
      journal = {\aap},
         year = 2015,
        month = feb,
       volume = {574},
          eid = {A83},
        pages = {A83},
          doi = {10.1051/0004-6361/201425222},
archivePrefix = {arXiv},
       eprint = {1412.3593},
 primaryClass = {astro-ph.EP},
       adsurl = {https://ui.adsabs.harvard.edu/abs/2015A&A...574A..83K}
}

@ARTICLE{Shima2009,
       author = {{Shima}, S. and {Kusano}, K. and {Kawano}, A. and {Sugiyama}, T. and {Kawahara}, S.},
        title = "{The super-droplet method for the numerical simulation of clouds and precipitation: a particle-based and probabilistic microphysics model coupled with a non-hydrostatic model}",
      journal = {Quarterly Journal of the Royal Meteorological Society},
         year = 2009,
        month = jul,
       volume = {135},
       number = {642},
        pages = {1307-1320},
          doi = {10.1002/qj.441},
archivePrefix = {arXiv},
       eprint = {physics/0701103},
 primaryClass = {physics.ao-ph},
       adsurl = {https://ui.adsabs.harvard.edu/abs/2009QJRMS.135.1307S}
}

@ARTICLE{Gundlach2015,
       author = {{Gundlach}, B. and {Blum}, J.},
        title = "{The Stickiness of Micrometer-sized Water-ice Particles}",
      journal = {\apj},
         year = 2015,
        month = jan,
       volume = {798},
       number = {1},
          eid = {34},
        pages = {34},
          doi = {10.1088/0004-637X/798/1/34},
archivePrefix = {arXiv},
       eprint = {1410.7199},
 primaryClass = {astro-ph.EP},
       adsurl = {https://ui.adsabs.harvard.edu/abs/2015ApJ...798...34G}
}

@ARTICLE{Wada2013,
       author = {{Wada}, Koji and {Tanaka}, Hidekazu and {Okuzumi}, Satoshi and {Kobayashi}, Hiroshi and {Suyama}, Toru and {Kimura}, Hiroshi and {Yamamoto}, Tetsuo},
        title = "{Growth efficiency of dust aggregates through collisions with high mass ratios}",
      journal = {\aap},
         year = 2013,
        month = nov,
       volume = {559},
          eid = {A62},
        pages = {A62},
          doi = {10.1051/0004-6361/201322259},
       adsurl = {https://ui.adsabs.harvard.edu/abs/2013A&A...559A..62W}
}

@ARTICLE{Musiolik2019,
       author = {{Musiolik}, Grzegorz and {Wurm}, Gerhard},
        title = "{Contacts of Water Ice in Protoplanetary Disks{\textemdash}Laboratory Experiments}",
      journal = {\apj},
         year = 2019,
        month = mar,
       volume = {873},
       number = {1},
          eid = {58},
        pages = {58},
          doi = {10.3847/1538-4357/ab0428},
archivePrefix = {arXiv},
       eprint = {1902.08503},
 primaryClass = {astro-ph.EP},
       adsurl = {https://ui.adsabs.harvard.edu/abs/2019ApJ...873...58M}
}

@ARTICLE{Krijt2024,
       author = {{Krijt}, Sebastiaan and {Arakawa}, Sota and {Oosterloo}, Mark and {Tanaka}, Hidekazu},
        title = "{A closer look at individual collisions of dust aggregates: material mixing and exchange on microscopic scales}",
      journal = {\mnras},
         year = 2024,
        month = nov,
       volume = {534},
       number = {3},
        pages = {2125-2133},
          doi = {10.1093/mnras/stae2247},
archivePrefix = {arXiv},
       eprint = {2409.17654},
 primaryClass = {astro-ph.EP},
       adsurl = {https://ui.adsabs.harvard.edu/abs/2024MNRAS.534.2125K}
}

@ARTICLE{Birnstiel2011,
       author = {{Birnstiel}, T. and {Ormel}, C.~W. and {Dullemond}, C.~P.},
        title = "{Dust size distributions in coagulation/fragmentation equilibrium: numerical solutions and analytical fits}",
      journal = {\aap},
         year = 2011,
        month = jan,
       volume = {525},
          eid = {A11},
        pages = {A11},
          doi = {10.1051/0004-6361/201015228},
archivePrefix = {arXiv},
       eprint = {1009.3011},
 primaryClass = {astro-ph.EP},
       adsurl = {https://ui.adsabs.harvard.edu/abs/2011A&A...525A..11B}
}

@ARTICLE{Gundlach2018,
       author = {{Gundlach}, B. and {Schmidt}, K.~P. and {Kreuzig}, C. and {Bischoff}, D. and {Rezaei}, F. and {Kothe}, S. and {Blum}, J. and {Grzesik}, B. and {Stoll}, E.},
        title = "{The tensile strength of ice and dust aggregates and its dependence on particle properties}",
      journal = {\mnras},
         year = 2018,
        month = sep,
       volume = {479},
       number = {1},
        pages = {1273-1277},
          doi = {10.1093/mnras/sty1550},
       adsurl = {https://ui.adsabs.harvard.edu/abs/2018MNRAS.479.1273G}
}

@ARTICLE{Hasegawa2023,
       author = {{Hasegawa}, Yukihiko and {Suzuki}, Takeru K. and {Tanaka}, Hidekazu and {Kobayashi}, Hiroshi and {Wada}, Koji},
        title = "{Collisional Growth and Fragmentation of Dust Aggregates. II. Mass Distribution of Icy Fragments}",
      journal = {\apj},
         year = 2023,
        month = feb,
       volume = {944},
       number = {1},
          eid = {38},
        pages = {38},
          doi = {10.3847/1538-4357/acadda},
archivePrefix = {arXiv},
       eprint = {2212.10796},
 primaryClass = {astro-ph.EP},
       adsurl = {https://ui.adsabs.harvard.edu/abs/2023ApJ...944...38H}
}

@ARTICLE{Supulver2000,
       author = {{Supulver}, Kimberley D. and {Lin}, D.~N.~C.},
        title = "{Formation of Icy Planetesimals in a Turbulent Solar Nebula}",
      journal = {\icarus},
         year = 2000,
        month = aug,
       volume = {146},
       number = {2},
        pages = {525-540},
          doi = {10.1006/icar.2000.6418},
       adsurl = {https://ui.adsabs.harvard.edu/abs/2000Icar..146..525S}
}

@ARTICLE{Hasegawa2021,
       author = {{Hasegawa}, Yukihiko and {Suzuki}, Takeru K. and {Tanaka}, Hidekazu and {Kobayashi}, Hiroshi and {Wada}, Koji},
        title = "{Collisional Growth and Fragmentation of Dust Aggregates with Low Mass Ratios. I. Critical Collision Velocity for Water Ice}",
      journal = {\apj},
         year = 2021,
        month = jul,
       volume = {915},
       number = {1},
          eid = {22},
        pages = {22},
          doi = {10.3847/1538-4357/abf6cf},
archivePrefix = {arXiv},
       eprint = {2104.06711},
 primaryClass = {astro-ph.EP},
       adsurl = {https://ui.adsabs.harvard.edu/abs/2021ApJ...915...22H}
}

@ARTICLE{Drazkowska2014,
       author = {{Dr{\k{a}}{\.z}kowska}, J. and {Dullemond}, C.~P.},
        title = "{Can dust coagulation trigger streaming instability?}",
      journal = {\aap},
         year = 2014,
        month = dec,
       volume = {572},
          eid = {A78},
        pages = {A78},
          doi = {10.1051/0004-6361/201424809},
archivePrefix = {arXiv},
       eprint = {1410.3832},
 primaryClass = {astro-ph.EP},
       adsurl = {https://ui.adsabs.harvard.edu/abs/2014A&A...572A..78D}
}

@ARTICLE{Ros2019,
       author = {{Ros}, Katrin and {Johansen}, Anders and {Riipinen}, Ilona and {Schlesinger}, Daniel},
        title = "{Effect of nucleation on icy pebble growth in protoplanetary discs}",
      journal = {\aap},
         year = 2019,
        month = sep,
       volume = {629},
          eid = {A65},
        pages = {A65},
          doi = {10.1051/0004-6361/201834331},
archivePrefix = {arXiv},
       eprint = {1907.08471},
 primaryClass = {astro-ph.EP},
       adsurl = {https://ui.adsabs.harvard.edu/abs/2019A&A...629A..65R}
}

@ARTICLE{Ros2024,
       author = {{Ros}, Katrin and {Johansen}, Anders},
        title = "{Fast formation of large ice pebbles after FU Orionis outbursts}",
      journal = {\aap},
         year = 2024,
        month = jun,
       volume = {686},
          eid = {A237},
        pages = {A237},
          doi = {10.1051/0004-6361/202348101},
archivePrefix = {arXiv},
       eprint = {2405.09237},
 primaryClass = {astro-ph.EP},
       adsurl = {https://ui.adsabs.harvard.edu/abs/2024A&A...686A.237R}
}

@ARTICLE{Youdin2007,
       author = {{Youdin}, Andrew N. and {Lithwick}, Yoram},
        title = "{Particle stirring in turbulent gas disks: Including orbital oscillations}",
      journal = {\icarus},
         year = 2007,
        month = dec,
       volume = {192},
       number = {2},
        pages = {588-604},
          doi = {10.1016/j.icarus.2007.07.012},
archivePrefix = {arXiv},
       eprint = {0707.2975},
 primaryClass = {astro-ph},
       adsurl = {https://ui.adsabs.harvard.edu/abs/2007Icar..192..588Y}
}

@ARTICLE{Gurrutxaga2026,
       author = {{Gurrutxaga}, Nerea and {Drazkowska}, Joanna and {Vaikundaraman}, Vignesh and {Kleine}, Thorsten},
        title = "{Carbonaceous Chondrites provide evidence for late-stage planetesimal formation in a pressure bump}",
      journal = {arXiv e-prints},
         year = 2026,
        month = apr,
          eid = {arXiv:2604.16604},
        pages = {arXiv:2604.16604},
archivePrefix = {arXiv},
       eprint = {2604.16604},
 primaryClass = {astro-ph.EP},
       adsurl = {https://ui.adsabs.harvard.edu/abs/2026arXiv260416604G}
}

\end{document}